\documentclass[%
superscriptaddress,
groupedaddress,
nofootinbib,
 amsmath,amssymb,
 aps,
prb,
 twocolumn,
 longbibliography
]{revtex4-1}

\usepackage{amsmath,amssymb,amsfonts,amsthm,lineno}
\usepackage{comment}
\usepackage{enumerate}
\usepackage{mathrsfs}
\usepackage{bbold}
\usepackage{tikz}
\usetikzlibrary{shapes.geometric, arrows}
\usepackage{hyperref}
\usepackage{slashed}
\usepackage{color}
\usepackage{empheq}
\usepackage{soul}
\usepackage{cancel}
\usepackage{placeins}
\usepackage{xargs}
\usepackage{tikz-cd}
\usepackage[caption=false]{subfig}

\makeatletter
\newcommand{\Vast}{\bBigg@{4.75}}
\makeatother

\newcommand{\be}{\begin{equation}}
\newcommand{\ee}{\end{equation}}
\newcommand{\bea}{\begin{eqnarray}}
\newcommand{\eea}{\end{eqnarray}}

\newcommand{\Nbar}{\overline{N}}

\makeatletter
\renewcommand{\@seccntformat}[1]{\csname the#1\endcsname.\,\,}
\makeatother
\allowdisplaybreaks

\usetikzlibrary{shapes,arrows,chains}
\usetikzlibrary{decorations.markings}
\usetikzlibrary{decorations.pathmorphing}
\usetikzlibrary{shapes.multipart}
\tikzset{snake it/.style={decorate, decoration=snake}}

\usepackage{tikz-cd}
\usetikzlibrary{cd,decorations.markings, arrows.meta}

\newcommand\qt\tau

\let \savenumberline \numberline
\def \numberline#1{\savenumberline{#1.}}

\makeatletter
\def\@fpheader{\relax}
\makeatother

\def\bea{\begin{eqnarray}}
\def\eea{\end{eqnarray}}

\begin{document}

\title{
Exact Regions of Superradiant Instability of \\ Kerr-Newman Black Holes and Massive Scalar Fields
}

\author{John Adrian B. Baybay}
    \email{jbaybay@nip.upd.edu.ph}
\author{Kevin T. Grosvenor}
    \email{kgrosvenor@nip.upd.edu.ph}
\affiliation{National Institute of Physics, University of the Philippines \\
Diliman, Quezon City 1101, Philippines}

\date{\today}

\begin{abstract}
We investigate the superradiant instability of Kerr–Newman black holes in the presence of a massive, charged scalar field using the Vieira–Bezerra–Kokkotas (VBK) method. We study the solutions of the exact polynomial condition for quasibound state frequencies and determine the domain of superradiant instability in parameter space without relying on the hydrogenic approximation or numerics. We derive the minimum scalar mass needed for quasibound states to exist, and identify the precise overlap region between the quasibound and superradiant conditions where instability can occur. We obtain perturbative and exact analytic expressions for the instability boundaries and growth rates, and clarify their relation to previous numerical results. Our analysis reveals how the instability region shifts from nearly neutral Kerr black holes for light fields to highly charged near-extremal Kerr–Newman black holes for heavier fields, while remaining absent in the Reissner–Nordstr\"om limit. 
\end{abstract}

\maketitle

\tableofcontents

\section{Introduction}

Since black holes are immersed in an environment, it is of great interest to investigate the effects on the black hole of its interaction with that environment (e.g. for gravitational wave searches \cite{cole2023distinguishing,cardoso2022black}, for the study of active galactic nuclei \cite{sarmah2025effects}, for black hole imaging \cite{lupsasca2024beginner}, jets \cite{oei2024black}, and for dark matter searches \cite{brito2025black, blas2022black, hui2023black}). These effects can present themselves as perturbations that may grow in time and eventually lead to the instability of the black hole. An example of such an instability is black hole superradiance. The effect of superradiance, as introduced by Dicke \cite{brito2020superradiance}, is a radiation enhancement process that can arise in systems with dissipation. This phenomenon can be observed in many different systems of interest in physics such as in optics, in quantum mechanics, in condensed matter systems, and especially in black holes (see \cite{bekenstein1998many} for a general review of superradiance as well as applications to different physical systems; see also \cite{lukin2022optical,sierra2022dicke,stransky2019superradiance,eleuch2014open} for various specific applications).

Black hole superradiance occurs when an incident field extracts energy away from the black hole resulting in an amplified scattered field. This can happen for modes with frequency $\omega$ below a certain critical frequency $\omega_c$, which depends on the spin and charge of the black hole. This effect can be further amplified by surrounding the black hole with hypothetical reflecting mirrors, thereby trapping the amplified field in the vicinity of the black hole and forcing it to undergo the superradiant process repeatedly. For as long as the superradiant condition is satisfied, the external field continues to extract energy from the black hole and its total energy grows exponentially until immense radiation pressure destroys the mirrors. This is known in the literature as a black hole superradiant instability, or a black hole bomb \cite{blackholebomb,cardoso2004black}. Though highly evocative, the hypothetical mirror setup is not, in fact, necessary to induce an instability of the black hole; one simply needs to consider massive fields to achieve the same effect. When the field has a mass $\mu$ such that $\mu > \omega$, the mass effectively acts as the reflecting mirror \cite{Damour:1976kh,ZOUROS1979139}, creating a cloud around the black hole. This is called a gravitational atom, as in a certain limit it behaves essentially analogously to the hydrogen atom \cite{baumann2019spectra,dias2023black,tomaselli2024legacy,tomaselli2024resonant,Baryakhtar:2020gao,baryakhtar2017black}. This phenomenon works for massive bosons and not for fermions due to the Pauli exclusion principle \cite{vicente2018penrose}. The approximations leading to the gravitational atom picture were first systematized by Starobinskii \cite{starobinskii1973amplification}, and later by Detweiler \cite{detweiler1980klein}, to solve the Klein-Gordon equation in a black hole background. Other natural approximations are the Wentzel-Kramers-Brillouin method and the eikonal limit \cite{Konoplya:2019hlu,Konoplya:2023moy}. Analytic results for quasinormal mode frequencies and grey-body factors can be derived using the method of continued fractions \cite{Suzuki:1998vy,Dolan:2007mj,Konoplya:2024vuj}.


It was then further posited that black holes can act as particle detectors which can be used as probes to detect new physics. One particular example of new physics one can hope to detect with the help of gravitational waves associated with black hole superradiant instability is the existence of ultralight bosons \cite{arvanitaki2015discovering}. Recent further developments include studies on the effects of gravitational atoms in binary systems \cite{baumann2022ionization}. 

In 2016, Vieira and Bezerra presented a new method to find exact solutions of the Klein-Gordon equation in different black-hole background spacetimes, including the Kerr-Newman black hole \cite{Vieira:2021xqw}. In 2021, Vieira and Kokkotas applied the same technique to study Schwarzschild acoustic black holes \cite{Vieira:2021xqw}. This was followed by a study of the superradiant instability of Kerr-Newman black holes in the presence of a massive scalar field using this method, which they then coined the Vieira–Bezerra–Kokkotas (VBK) method \cite{vieira2022instability}. This method uses polynomial conditions of the Heun function to solve for resonant modes that form quasibound states around the black hole. The core result of \cite{vieira2022instability} is to furnish a quartic equation, called the characteristic equation, for the resonant frequencies of the scalar field modes. The solutions to this quartic equation were plotted and to check for instabilities, the authors looked at the conditions under which the solution grows over time, that is, $\text{Im} \, \omega > 0$, where $\omega$ is the mode frequency. 

In this paper, we will use the VBK method presented in \cite{vieira2022instability} to completely specify the regions of black hole superradiant instability as a function of the parameters of the black hole (spin $a$ and charge $Q$) and the surrounding scalar field (mass $\mu$). There are plenty of studies of black hole superradiant instability for the Kerr black hole \cite{press1972floating,dolan2007instability,east2017superradiant}. The majority of these studies work in the limit of a very light external scalar field with a mass parameter $\mu$ satisfying $\mu M \ll 1$, where $M$ is the mass of the black hole. There are likely at least three reasons for this: 1. The lighter the field is, the broader is the range of $a$ over which superradiant instability can occur; 2. The smaller the value of $a$, the smaller the value of $\mu$ needed for superradiant instability; and 3. Most phenomenological models take the scalar field to be the QCD axion \cite{arvanitaki2011exploring,yoshino2012bosenova,baryakhtar2021black,xie2025self}, which is constrained to be extremely light. As such, there does not appear to be in the literature a complete and systematic characterization of the region of superradiant instability in the parameter space spanned by the spin $a$ and charge $Q$ of the Kerr-Newman black hole and the mass $\mu$ and charge $q$ of the scalar field. The goal of this paper is to fill in this gap. 

We will furnish a plot, in fig. \ref{subfig:instability} that shows the region of the space of Kerr-Newman black holes that are unstable to superradiance given the mass of the external scalar field. Compare this, for example, with \cite{Myung:2022biw}, which describes conditions for superradiant instability of Kerr-Newman black holes due to a charged massive scalar field in terms of the existence of a ``trapping'' effective potential, though, in truth, derives explicit results only in the Kerr black hole ($Q=0$) limit. The VBK method allows us to exactly characterize the regions of black-hole parameter space that are unstable to superradiance without the need for an intervening analysis of the effective potential by working directly with the modes of the scalar field themselves. This problem serves as an excellent testing ground for this method and provides a good opportunity to compare it with the methods of hydrogenic approximation and numerics.

The presentation of the paper will proceed as follows. In section \ref{sec:2}, we will introduce the Klein-Gordon equation in the Kerr-Newman background. In section \ref{sec:3}, we give a quick overview of the VBK method as well as the standard hydrogenic approximation. In section \ref{sec:4}, we will then show how to solve for the quasibound states and superradiant states and plot their respective regions in the parameter space. We take the overlap of these two regions to determine the region of instability. In section \ref{sec:5}, we fix $a$ and $Q$, instead tuning $q$ to determine the region where the scalar field mode grows over time and compare with the numerical results of Furuhashi and Nambu \cite{furuhashi2004instability}. We highlight the discrepancies between the numerical results and our exact analytical results and identify the origin of these discrepancies. Finally, we conclude with some discussions and future directions. 

Throughout this paper, we use the reduced units $G = \hbar = c =1$. In these units, the (gravitational) mass $M$ of the black hole has units of \emph{length}. However, the Klein-Gordon equation \eqref{eq:KGeqtn} implies that the (inertial) mass $\mu$ of the scalar field has units of \emph{inverse length}, since $\mu^2$ must have the same units as the d'Alembert operator. Similarly, $a$ and $Q$ have the same units as $M$, namely length (c.f. eq. \eqref{eq:rD}). However, $q$ has units of inverse length since the gauge covariant derivative is $\nabla + i qA$ and the gauge field $A$ is dimensionless.

We are always free to set $M=1$, which is equivalent to ``normalizing by $M$'', or writing expressions purely in terms of $M$ and the dimensionless quantities $\frac{a}{M}$, $\frac{Q}{M}$, $\mu M$, and $q M$. We may normalize by $M$ explicitly by retaining the explicit factors of $M$ or else we can normalize by $M$ implicitly by setting $M=1$. Typical of the literature, we do both at various points in this paper. No confusion should arise since there is a unique way to reintroduce $M$ after it has been set to 1. The only potential confusion may arise if the value of any of the remaining four parameters is fixed. For example, if we set $M=1$ and $qM = 1$, then the expression $Q$ is ambiguous as it may mean $qQ$ or $\frac{Q}{M}$. Therefore, we will state clearly when we do fix the values of any of the other parameters.


\section{Klein-Gordon Field in the Kerr-Newman Background}
\label{sec:2}

A stationary axisymmetric charged rotating black hole may be modeled using the Kerr Newman (KN) metric \cite{adamo2014kerr}. In the Boyer-Lindquist coordinates, the line element is
\begin{align} \label{eq:ogmetric}
    ds^2 &= \frac{\Delta}{\rho^2} \bigl( dt - a \sin^2 \theta \, d \phi )^2 - \frac{\rho^2}{\Delta}  dr^2 - \rho^2 d\theta^2 \notag \\
    &\quad - \frac{\sin^2 \theta}{\rho^2} \bigl[ (r^2 + a^2 ) d \phi - a dt \bigr]^2,
\end{align}
where
\begin{subequations} \label{eq2}
\begin{align}
	\Delta &= r^2 + a^2 - 2Mr + Q^2, \\
	 \rho^2 &= r^2 + a^2 \cos^2 \theta,
\end{align}
\end{subequations}
with $M$ being the mass of the black hole, $a$ its spin, and $Q$ its charge. The limit $a=0$ corresponds to the Reissner-Nordstr\"om (RN) black hole, the limit $Q=0$ corresponds to the Kerr black hole, and the limit $a = Q = 0$ corresponds to the Schwarzschild black hole. The KN metric has a physical singularity at $\rho = 0$. There are also coordinate singularities at the roots of the equation $\Delta=0$, which are 
\begin{align} \label{eq:rD}
r_\pm &= M \pm D, &%
D &= \sqrt{M^2-a^2-Q^2}.
\end{align}
Thus, we can write
\begin{equation}
    \label{eq:Delta}
    \Delta = (r- r_+) (r - r_-).
\end{equation}
The surface $r = r_+$ is called the outer horizon and $r=r_-$ the inner horizon. Lastly, the KN spacetime is endowed with a Maxwell potential given by
\begin{equation}
A_{\mu} = \frac{Q r}{\rho^2} (1, 0, 0, -a \sin^2 \theta ).    
\end{equation}

The equation of motion that describes a scalar field $\Psi$, with mass $\mu$ and charge $q$, propagating in a curved spacetime and in the presence of an electromagnetic field, is given by
\begin{equation} \label{eq:KGeqtn}
\bigl[ ( \nabla_{\nu} + i q A_{\nu} ) ( \nabla^{\nu} + iq A^{\nu} ) + \mu^2 \bigr] \Psi = 0.
\end{equation}
We assume a separable solution
\begin{equation} \label{eq:ansatz}
    \Psi(t, r, \theta, \phi) = e^{-i \omega t } R(r) P (\theta) e^{i m \phi },
\end{equation}
with $R(r)$ the radial function, $P(\theta)$ the angular function, and $m$ the magnetic quantum number. Eq. \eqref{eq:KGeqtn} separates into two ordinary differential equations given by
\begin{equation} \label{eq:KGangle}
    \frac{1}{\sin \theta} \frac{d}{d\theta} \bigg[\sin\theta \frac{d P}{d\theta}\bigg] + \Lambda_{\theta} P = 0,
\end{equation}
and
\begin{equation} \label{eq:KGradial}
    \Delta \frac{d}{dr}\bigg[\Delta \frac{d R}{d r} \bigg] + ( \Omega^2 - \Lambda_r \Delta ) R = 0, 
\end{equation}
where
\begin{subequations}
\begin{align}
\Lambda_{\theta} &= \lambda + a^2 (\mu^2 - \omega^2) \sin^2\theta - \frac{m^2}{\sin^2\theta}, \\
\Lambda_r &= \lambda + \mu^2(r^2 + a^2) - 2 a m \omega, \\
\Omega &= \omega (r^2 + a^2) - am - qQr,
\end{align}
\end{subequations}
and $\lambda$ is a separation constant. The angular equation may be solved using spheroidal functions \cite{Berti:2005gp} that reduce to the Legendre functions when $\omega = \mu$. From here it is standard to use numerical techniques to solve the radial equation. However, Starobinskii \cite{starobinskii1973amplification} showed that the equation can be solved analytically in the limit $\omega M \ll 1$ and $\mu M \ll 1$. In the region $r \gg M$, the equations reduce to those of the hydrogen atom, with $\lambda \approx \ell ( \ell + 1 )$ and $\ell$ being the azimuthal quantum number \cite{detweiler1980klein}. Recently, however, Viera, Bezerra, and Muniz \cite{vieira2022instability} used the VBK method to find exact solutions of eq. \eqref{eq:KGeqtn} involving the confluent Heun functions. For our purposes, the main result of this method is an exact equation, eq. \eqref{eq:resoeq1}, for the resonance frequency of modes of the charged scalar field in a Kerr-Newman black hole background. This can then be used to determine the conditions under which black hole superradiant instability will take place:
\begin{enumerate}[$\bullet$]
    \item The formation of bound states around the black hole requires $\text{Re} \bigl[ \sqrt{\mu^2 - \omega^2} \bigr] > 0$; and
    \item The superradiant condition $\omega < \omega_c$ must be satisfied.
\end{enumerate}


\section{VBK method vs. Hydrogenic Approximation}
\label{sec:3}

To determine quasibound states in the hydrogenic approximation we consider scalar fields that are light enough such that their Compton wavelength is much larger than that of the black hole. That is, defining $\alpha = \mu M$,  which can be thought of as the gravitational fine structure constant \cite{baumann2019spectra}, this limit is $\alpha \ll 1$. In this limit, the Klein-Gordon equation for a scalar field in the Kerr black hole background (we set $Q = 0$ for now for simplicity) takes the form of a Schr\"odinger equation, written as
\begin{equation}
    i\dfrac{\partial \Psi}{\partial t} = - \dfrac{1}{2 \mu } \nabla^2 \Psi + V_{\text{eff}}(r) \Psi,
    \label{eq:hydrolike}
\end{equation}
where the effective potential is 
\begin{equation}
    V_{\text{eff}} = - \dfrac{\alpha}{r} + \dfrac{\ell(\ell +1)}{2 \mu r^2}.
\end{equation}
The form of equation \eqref{eq:hydrolike} is reminiscent of the hydrogen atom and allows us to solve for the spectrum by analogy \cite{detweiler1980klein}:
\begin{align}
    E_n &= - \frac{\alpha^2 \mu}{2 n^2}, &%
    \text{Re} \, \omega_n = \mu + E_n = \mu \biggl( 1 - \frac{\alpha^2}{2 n^2} \biggr). 
\end{align}
The radial extent of the scalar field cloud is determined in a way analogous to the Bohr radius and orbital radii of hydrogen. The growth rate of the cloud is determined by the imaginary part of the frequency.

In contrast, the VBK method proceeds without approximations. The radial equation \eqref{eq:KGradial} reads
\begin{equation}
\begin{aligned}
0 &= \frac{d^2 R}{d r^2} + \left(\frac{1}{r-r_{+}}+\frac{1}{r-r_{-}}\right) \frac{d R}{d r} \\
&\quad + \biggl( \frac{\Omega^2}{\Delta^2} - \frac{\Lambda_r}{\Delta} \biggr) R.
\end{aligned}
\end{equation}
This form of the radial equation indicates that there are two finite regular singularities associated with the two event horizons, i.e. $r_+$ and $r_-$, and an irregular singularity at infinity. Thus, we can perform the M\"obius transformation
\begin{equation}
x = \frac{r-r_-}{r_+ - r_-}
\end{equation}
to move the regular singularity at $r = r_-$ to $x=0$, the one at $r=r_+$ to $x=1$, and the irregular singularity at $r = \infty$ to $x = \infty$. The homotopic transformation
\begin{equation}
R (r) = x^{D_0} (x-1)^{D_1} e^{D_2 x} U(x),
\end{equation}
where the constants $D_i$ are given by
\begin{subequations}
\begin{align}
D_0 &= -i \frac{\omega ( r_{-}^{2} + a^2 ) - am - qQ r_-}{r_+ - r_-}, \\
D_1 &= -i \frac{\omega ( r_{+}^{2} + a^2 ) - am - qQ r_+}{r_+ - r_-}, \\
D_2 &= - (r_+ - r_- ) \sqrt{\mu^2 - \omega^2},
\end{align}
\end{subequations}
turns the radial equation into the confluent Heun equation 
\begin{equation}
\begin{aligned}
0 &= \frac{d^2 U}{dx^2} + \biggl( 2D_2 + \frac{1 + 2D_0}{x} + \frac{1+2D_1}{x-1} \biggr) \frac{dU}{dx} \\
&\quad + \biggl( \frac{\xi}{x} + \frac{\zeta}{x-1} \biggr) U,
\end{aligned}
\end{equation}
where $\xi$ and $\zeta$ are complicated expressions in terms of the parameters $a$, $Q$, $\mu$, $q$, $M$, $m$, and $\omega$. The confluent Heun functions of first and second kind furnish solutions, with potentially different linear combinations of these being appropriate in a neighborhood of each of the three singularities. These solutions must be made to match along the intersection of their domains. The result is a polynomial condition similar in principle to the one that derives from the ``halting'' problem of the expansions involved in the Frobenius method (e.g., that the wave functions of the 1d quantum harmonic oscillator must take the form of a finite-degree polynomial multiplying a Gaussian -- the Hermite functions). The overlapping expansions around each singularity complicates matters, but the procedure is otherwise analogous. The degree of the polynomial that we must force the confluent Heun functions to become is called the overtone number, $N$, which must thus be a non-negative integer. In the same way that the halting problem in the Frobenius method determines the spectrum (e.g., by expressing the harmonic oscillator energy in terms of the finite degree $n$ of the associated Hermite polynomial), the VBK method furnishes an exact equation for the resonance frequencies $\omega$ of the modes of the scalar field in terms of the overtone number $N$, the black hole parameters $a$, $Q$, and $M$, and the scalar field parameters $\mu$ and $q$. This can be written as a quartic equation for $\omega$. We present this equation in a way that is far more suitable for analytic manipulation than is the original presentation in \cite{vieira2022instability}. First, we define
\begin{subequations}
\begin{align}
\Nbar &= N+1, \\
A &= am + MQq, \\
B &= 2M^2 - Q^2,
\end{align}
\end{subequations}
which we use to define
\begin{subequations}
\begin{align}
b_2 &= A^2 - (Q^2 q^2 + \overline{N}^2 ) D^2, \\
b_3 &= 2 (2MQqD^2 - AB), \\
c_2 &= -2A \Nbar D, \\
c_3 &= 2B \Nbar D.
\end{align}
\end{subequations}
Finally, the characteristic resonance equation reads
\begin{equation} \label{eq:resoeq1}
a_0 + \sum_{n=1}^{4} a_n ( \omega^n - \mu^n ) = 0,
\end{equation}
where
\begin{subequations}
\begin{align}
a_0 &= - (M \mu - Qq)^2 D^2 \mu^2, \\
a_1 &= (2MQq D^2 - a_3 ) \mu^2, \\
a_2 &= b_2 + ic_2 - a_4 \mu^2, \\
a_3 &= b_3 + ic_3, \\
a_4 &= B^2 - 4M^2 D^2.
\end{align}
\end{subequations}
Note that the coefficients $a_n$ above coincide with those in \cite{vieira2022instability} except for $a_0$. There are generally four solutions for $\omega$ and, in principle, there is a closed-form formula for these involving only arithmetic operations and roots. However, this formula is complicated and cumbersome to use and when applying any approximations or limits, it is generally simpler and more transparent to apply them to the equation first rather than directly on the solutions themselves. Furthermore, we are only interested in the solution that satisfies $\omega \approx \mu$, but which one that is out of the four solutions changes depending on the various parameters due to branch ambiguities in the definitions of the root operations. This problem is well documented in \cite{vieira2022instability}. We now proceed to describe how we will use and analyze this equation.


\section{The Regions of Superradiant Instability}
\label{sec:4}

We will now examine the conditions that must be satisfied in order for the black hole to develop a superradiant instability. We require that there be both quasibound states and superradiant modes. The existence of quasibound states allows the formation of scalar field clouds surrounding the black hole that can continuously extract energy from the black hole via superradiant scattering. 


\subsection{Quasibound States}

\begin{figure*}[t!]
     \centering
     \subfloat[Kerr ($Q=0$)]{%
         \includegraphics[width= 0.47 \textwidth]{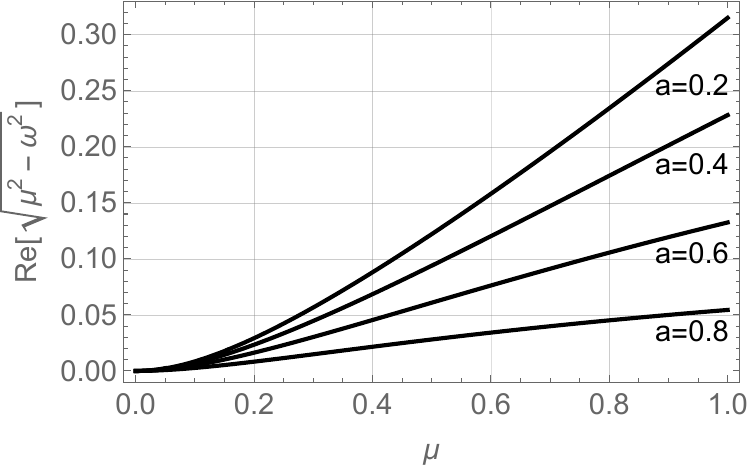}
         \label{fig:ReKerr}%
         }
     \hfill
     \subfloat[Reissner-Nordstr\"om ($a=0$)]{%
         \includegraphics[width= 0.47 \textwidth]{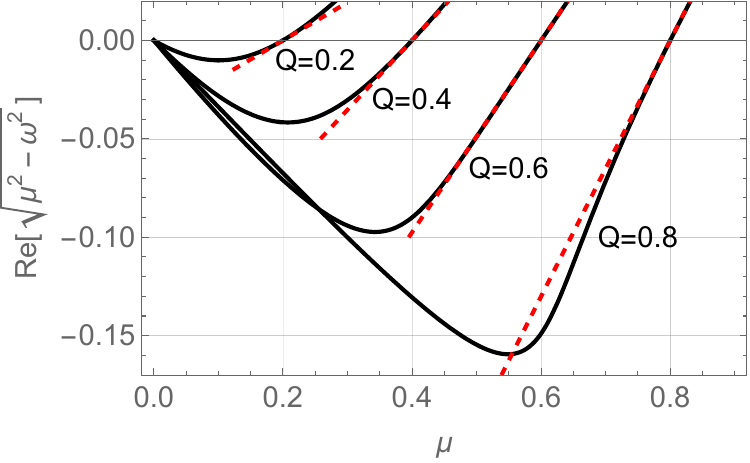}
         \label{fig:ReRN}%
         }
     \hfill
     \subfloat[$a=0.1$, $Q = 0.5$]{%
         \includegraphics[width= 0.47 \textwidth]{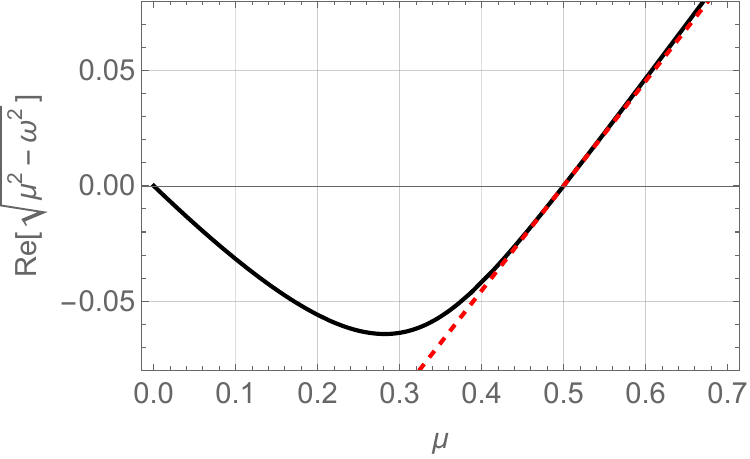}
         \label{fig:Re_ap1_Qp5}%
         }
     \hfill
     \subfloat[$a=0.5$, $Q=0.5$]{%
         \includegraphics[width= 0.47 \textwidth]{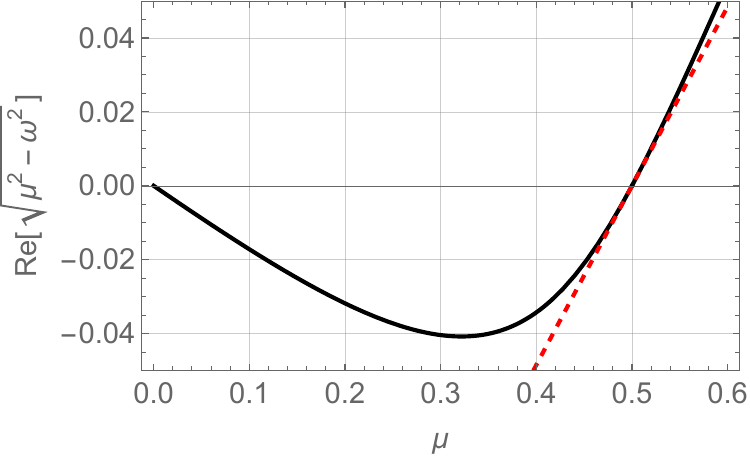}
         \label{fig:Re_ap5_Qp5}%
         }
       \hfill
     \subfloat[$a=0.7$, $Q=0.1$]{%
         \includegraphics[width = 0.5 \textwidth]{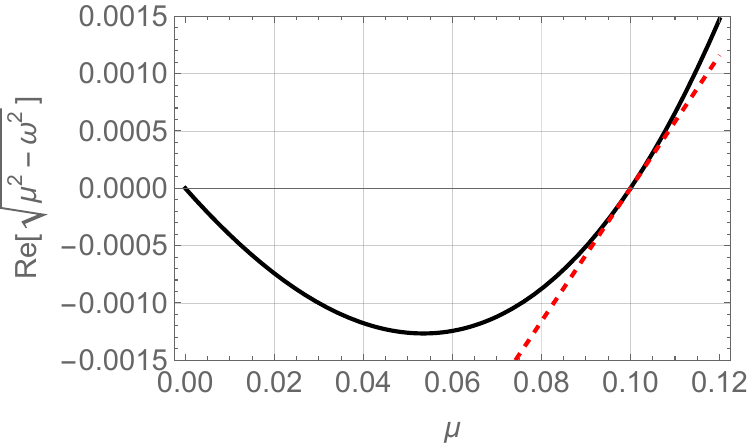}
         \label{fig:Re_ap7_Qp1}%
         }
        \hfill
     \subfloat[$a=0.1$, $Q=0.7$]{%
         \includegraphics[width = 0.47 \textwidth]{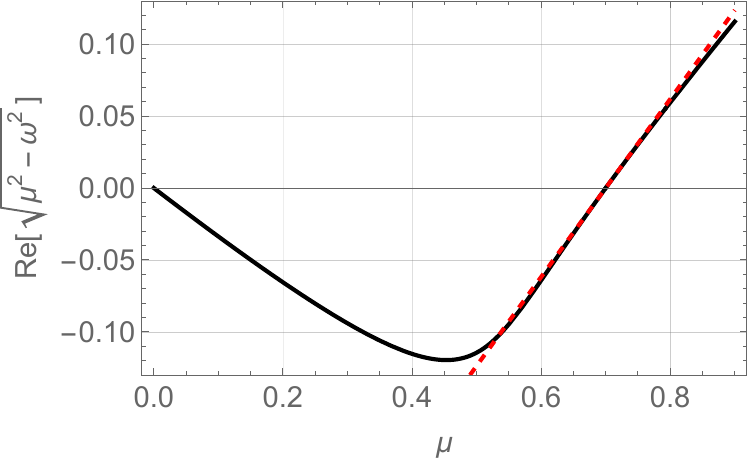}
         \label{fig:Re_ap1_Qp7}%
         }
        \caption{Plots of $\text{Re} \bigl[ \sqrt{\mu^2 - \omega^2} \bigr]$ versus $\mu$ for the (a) Kerr black hole with various different values of $a$, (b) Reissner-Nordstr\"om black hole with various different values of $Q$, and for (c)-(f) Kerr-Newman black hole with different values of $a$ and $Q$. The linear approximation \eqref{eq:Reslope} at the nontrivial root is plotted as red dashed lines. For the Kerr black hole, the plots vanish exactly at $\mu = 0$ and are flat, as predicted by \eqref{eq:Reslope}.}
        \label{fig:ReKN}
\end{figure*}

To determine if there are quasibound state modes, we must first solve the radial equation and then impose the boundary conditions
\begin{enumerate}[1.]
\item The radial solution should describe an incoming wave at the outer event horizon; 
\item The radial solution must tend to zero far from the black hole at asymptotic infinity.
\end{enumerate} 
Let $R_1$ denote the expansion of the the radial function near $r=r_+$. The first condition is satisfied by setting the solution to be
\begin{equation}
R_1 (r) \sim C_1 (r-r_+) e^{- \frac{i}{2 \kappa_+} ( \omega - \omega_c )},
\end{equation}
where $C_1$ is a constant,
\begin{equation}
\kappa_+ = \frac{r_+ - r_-}{2 ( r_{+}^{2} + a^2 )},
\end{equation}
and the critical frequency is given by
\begin{equation}
\omega_c = m \Omega_+ + q \Phi_+,
\end{equation}
where
\begin{align}
\Omega_+ &= \frac{a}{r_{+}^{2} + a^2}, &%
\Phi_+ &= \frac{Q r_+}{r_{+}^{2} + a^2}.
\end{align}
Let $R_{\infty}$ be the asymptotic form of the solution as $r \rightarrow \infty$. Then, we find
\begin{equation}
R_{\infty} (r) \sim C_{\infty} r^{- ( 1 + p )} e^{- \sqrt{\mu^2 - \omega^2} \, r},
\end{equation}
where
\begin{equation}
    p = \frac{q Q \omega + M (\mu^2 - 2 \omega^2)}{\sqrt{\mu^2 - \omega^2}}.
\end{equation}
The asymptotic boundary condition requires that 
\begin{equation}
\text{Re} \bigl[ \sqrt{\mu^2 - \omega^2} \bigr] > 0.
\end{equation}
Therefore, $\omega = \mu$ at the boundary of the quasibound-state region, in which case the equation reduces to
\begin{equation}
(M \mu - Qq)^2 D^2 \mu^2 = 0,
\end{equation}
the nontrivial solution to which is
\begin{equation}
\mu = \mu_0 \equiv \frac{Qq}{M}.
\end{equation}
We expand \eqref{eq:resoeq1} around $\mu = \mu_0$ and solve for $\omega$ to lowest order in $\mu - \mu_0$, giving the result
\begin{equation} \label{eq:Reslope}
\text{Re} \bigl[ \sqrt{\mu^2 - \omega^2} \bigr] = s ( \mu - \mu_0 ),
\end{equation}
where the slope is given by
\begin{align}
s &= \frac{\overline{N} D^2 \mu_0 M}{( \overline{N} D )^2 + C^2}, &%
C &= am + \mu_0 (Q^2 - M^2 ).
\end{align}
Note that $\text{Re} \bigl[ \sqrt{\mu^2 - \omega^2} \bigr]$ is positive for $\mu > \mu_0$ and negative for $\mu < \mu_0$. This is perfectly clear in fig. \ref{fig:ReKN}, except, of course, that both $\mu_0$ and the linear approximation \eqref{eq:Reslope} vanish identically for the Kerr black hole, consistent with the fact that the Kerr black hole plots are flat at $\mu = 0$. This analysis holds in the vicinity of $\mu = \mu_0$, but the plots show this to hold beyond this point up to and even beyond the $\mu M = 1$ threshold. Therefore, there are no quasibound states when $\mu < \mu_0$ and there are quasibound states when $\mu > \mu_0$. Furthermore, the linear approximation \eqref{eq:Reslope} faithfully reproduces the curves near the nontrivial root.

%


\subsection{Superradiant Modes}

\begin{figure*}[t!]
        \subfloat[Kerr]{%
            \includegraphics[width=.41\linewidth]{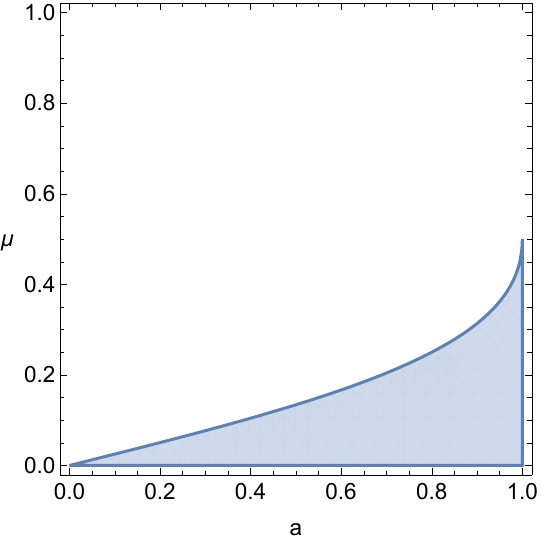}%
            \label{subfig:SR Kerr}%
        }\hfill
        \subfloat[Reissner-Nordstr\"om]{%
            \includegraphics[width=.41\linewidth]{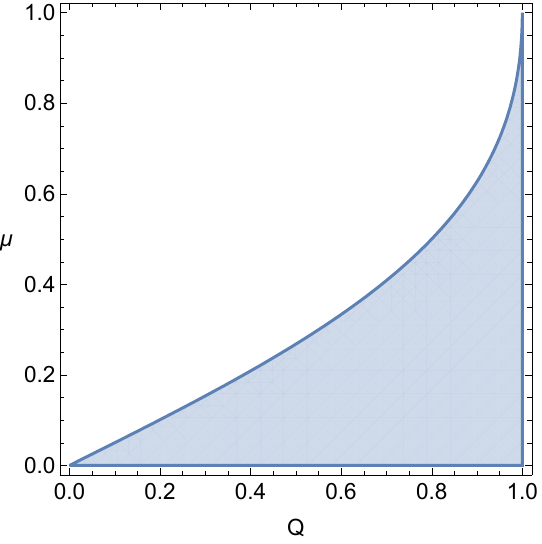} \hspace{1.1cm}%
            \label{subfig:SR RN}%
        }\hfill
         \subfloat[Kerr-Newman $\mu$ vs. $a$]{%
            \includegraphics[width=.49\linewidth]{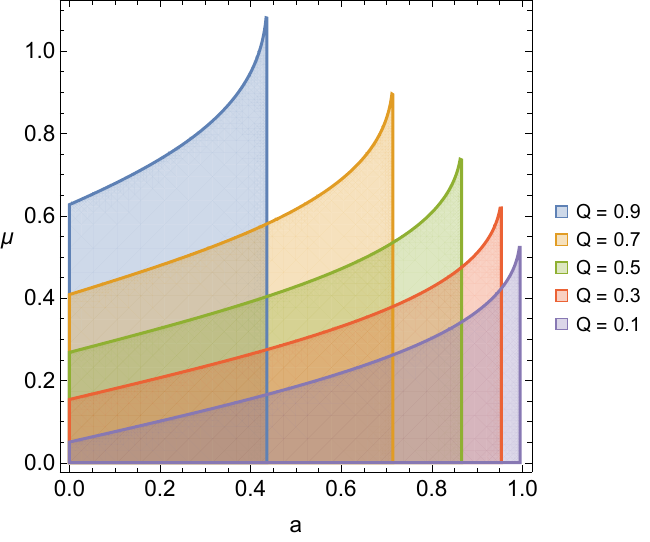}%
            \label{subfig:SR_KN_Qs}%
        }\hfill
        \subfloat[Kerr-Newman $\mu$ vs. $Q$]{%
            \includegraphics[width=.49\linewidth]{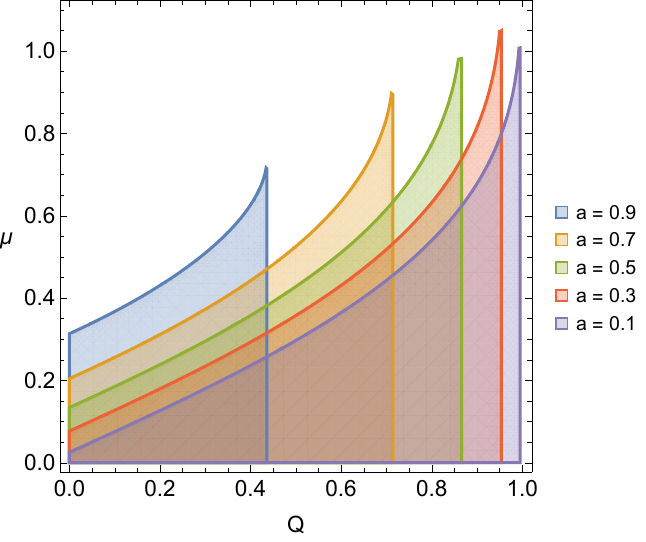}%
            \label{subfig:SR_KN_as}%
        }\hfill
        \caption{Plots of the superradiant region for the (a) Kerr black hole, (b) Reissner-Nordstr\"om black hole, (c) Kerr-Newman black hole as a function of $a$ for various values of $Q$, and the (d) Kerr-Newman black hole as a function of $Q$ for various values of $a$.}
    \label{fig:SR plots}
\end{figure*}

The critical frequency $\omega_c$ below which superradiance occurs is \cite{brito2020superradiance},
\begin{equation}
     \omega < \omega_c = \frac{m a + q Q r_+}{r_+^2 + a^2}.
\end{equation}
Since $\omega \lesssim \mu$, we can approximate the condition that $\mu$ must satisfy to allow for superradiance to be
\begin{equation} \label{eq:SRcond}
     \mu < \omega_c = \frac{m a + q Q r_+}{r_+^2 + a^2}.
\end{equation}
We plot this region for the Kerr black hole in fig. \ref{subfig:SR Kerr}, for the Reissner-Nordstr\"om black hole in fig. \ref{subfig:SR RN}, and for the Kerr-Newman black hole with $\mu$ vs. $a$ for various values of $Q$ in fig. \ref{subfig:SR_KN_Qs} and $\mu$ vs. $Q$ for various values of $a$ in fig. \ref{subfig:SR_KN_as}.

\begin{figure*}[t!]
        \subfloat[The superradiance region, quasibound state region, and their overlap for $\mu = 0.2$.]{%
            \includegraphics[width=.444\linewidth]{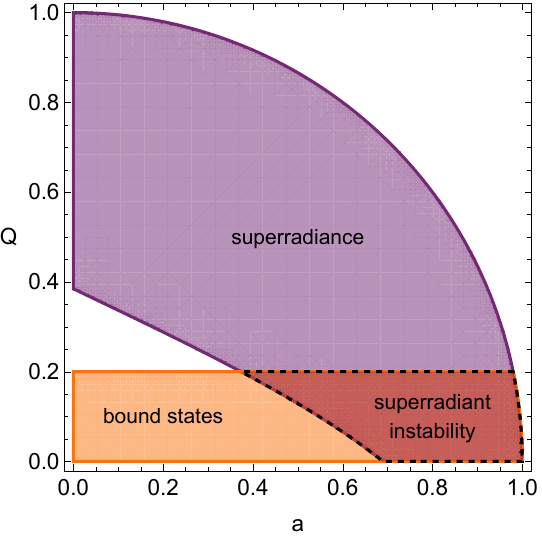}%
            \label{subfig:overlap}%
        }\hfill
        \subfloat[Superradiant instability regions for different values of scalar field mass $\mu$.]{%
            \includegraphics[width=.52\linewidth]{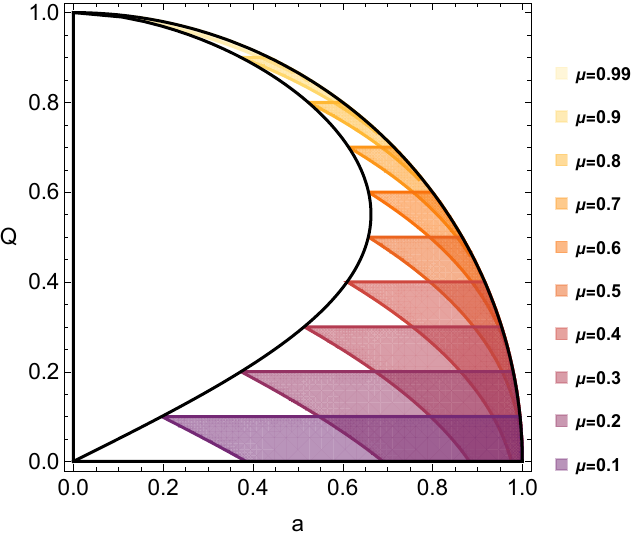}%
            \label{subfig:instability}%
        }\hfill
        \caption{Regions of superradiant instability with $M = 1,\, q = 1,\, m =1, \, N=0$.}
    \label{fig:paraplot}
\end{figure*}

It is particularly instructive to plot the regions of superradiance in the plane of black hole parameters $a$ and $Q$ for various different values of the scalar field mass parameter $\mu$. Furthermore, we can superpose on this the region where there are quasibound states. For these plots, we set $M = 1$, $m=1$, and $q=1$. Thus, the quasibound state region is simply the region $Q < \mu$. The superradiant region is simply given by \eqref{eq:SRcond}. For example, these regions are shown in fig. \ref{subfig:overlap} for the value $\mu = 0.2$. We are interested in the overlap between these two regions, of course, which is the region where the black hole may develop a superradiant instability. The region of superradiant instability is shown in fig. \ref{subfig:instability} for various values of $\mu$. The boundary of the union of the superradiant instability regions for all values of $\mu$ from 0 to 1 is the solid black curve in fig. \ref{subfig:instability} and is given by setting $\mu = Q$ in \eqref{eq:SRcond} and solving for $a$. Since this expression is genuinely confusing if we set both $M=1$ and $q=1$, we will write the general expression out in full detail, even including $m$ explicitly:
\begin{equation}
\bar{a} > |qQ| \frac{m (1 - \bar{Q}^2 ) + \sqrt{(1 - \bar{Q}^2 ) [ m^2 + (qQ)^2 \bar{Q}^2 ]}}{m^2 + (qQ)^2},
\end{equation}
where $\bar{a} = \frac{a}{M}$ and $\bar{Q} = \frac{Q}{M}$. Since we must also satisfy the condition $\bar{a}^2 + \bar{Q}^2 \leq 1$, the region shrinks down to two points in the RN limit: $\bar{Q} = 0$ and $\bar{Q}=1$. This is consistent with the well-known fact that the Reissner-Nordstr\"om black hole is stable against superradiance \cite{moncrief1974stability,hod2013no,hod2015stability}. 

The plot also confirms that, for very small values of the scalar field mass parameter $\mu$, the region of superradiant instability is confined to equally small values of the black hole charge $Q$. At values of $\mu$ appropriate for the QCD axion, we would essentially be confined to the Kerr black hole case. However, if we were to entertain models with much larger values of $\mu$, then we can see that a significant region opens up with considerable black hole charge. In fact, the extent of the superradiant instability region along the Kerr axis diminishes as $\mu$ increases and even vanishes for $\mu > 0.5$. Thus, above this value of the scalar field mass, we would only encounter superradiant instability for a Kerr-Newman black hole with some nonzero charge. As we continue to increase the scalar field mass, the black hole would have to become more charged and closer to extremal in order to potentially be unstable to superradiance.

\section{Scalar Field Growth Rate}
\label{sec:5}

\begin{figure*}[t!]
     \centering
         \includegraphics[width= 0.49\textwidth]{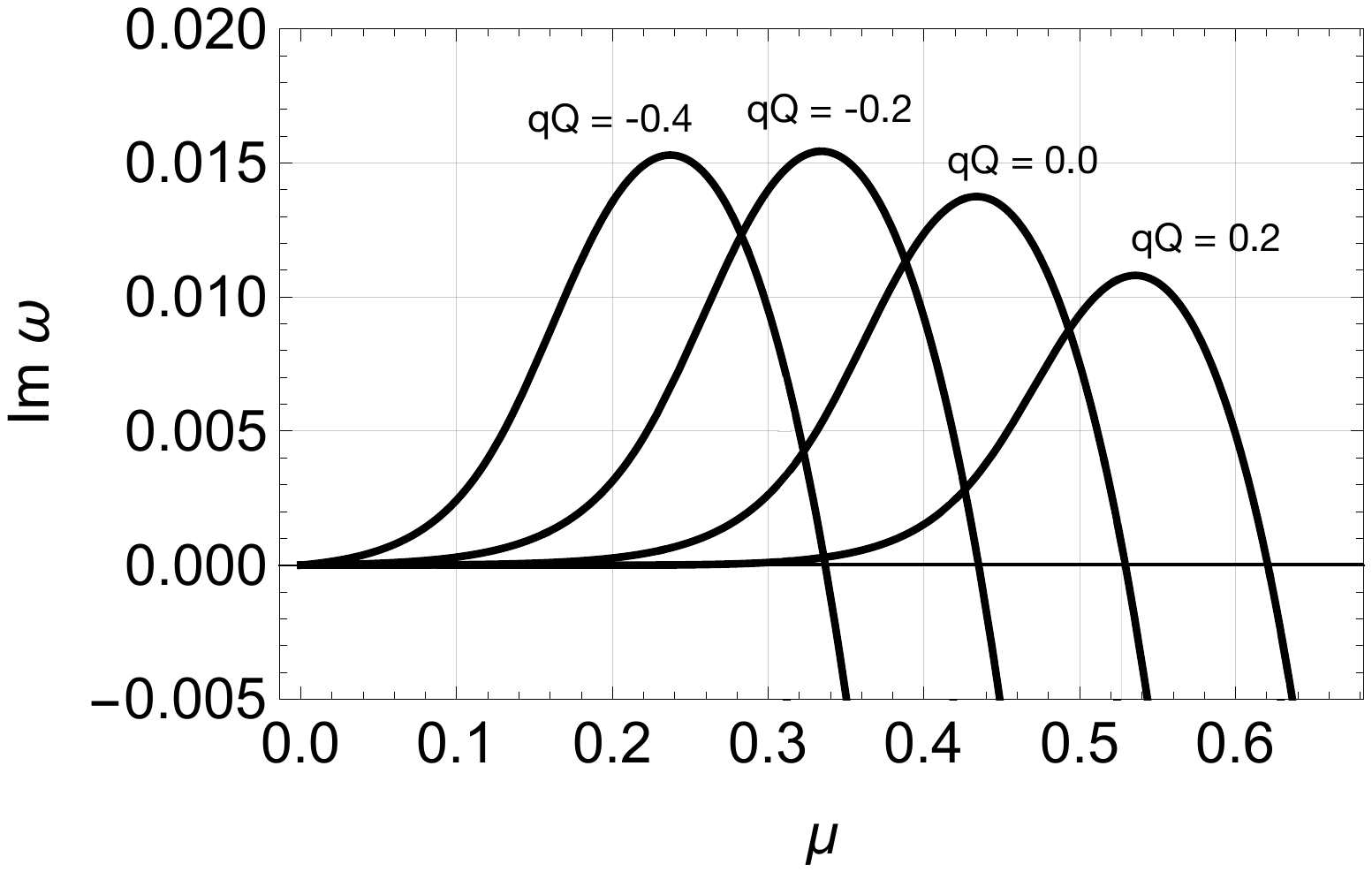}
     \hfill
         \includegraphics[width= 0.49\textwidth]{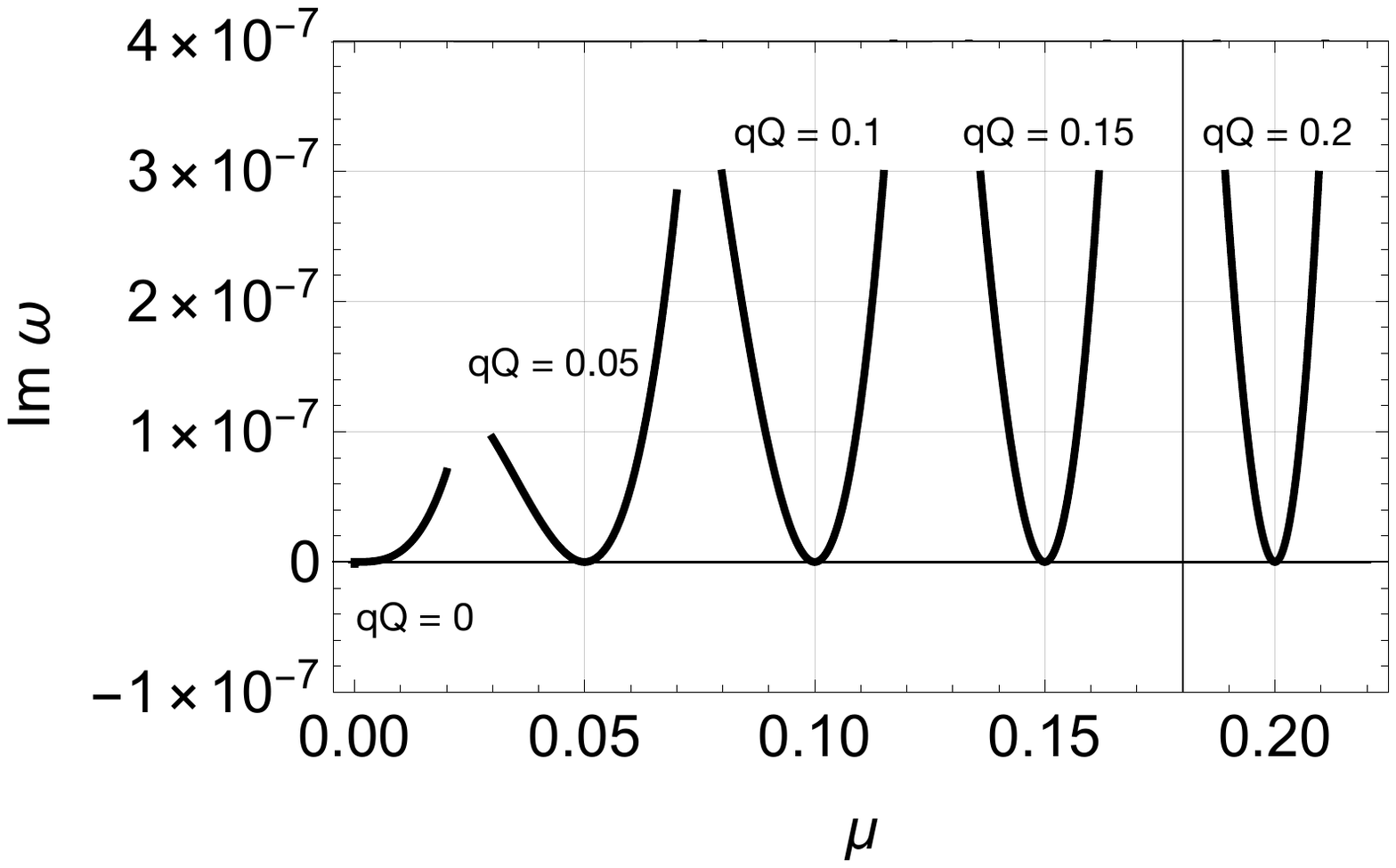}
         \vspace{-0.6cm}
        \caption{Plots of $\text{Im}\, \omega$ vs. $\mu$ with $a=0.98 M$, $Q=0.01M$, $M = 1$, $m=1$, $N = 0$ and $q$ as the tuning parameter. The left figure shows the points where $\text{Im} \, \omega$ turns negative for different values of $qQ$. The right figure shows portions of the plots zoomed into the region near $\mu = \mu_0 = qQ/M$ for several positive values of $qQ$.}
        \label{fig:imaginary_omega}
\end{figure*}

For the instability to occur effectively, the scalar field mode must grow over time. The growth rate, $\Gamma$, is
\begin{equation}
\Gamma = \text{Im} ( \omega ).
\end{equation}
Therefore, for the scalar field mode to grow, we must impose the condition
\begin{equation}
\text{Im}\, \omega > 0.
\end{equation}
Expanding eq. \eqref{eq:resoeq1} around $\mu = \mu_0$ and solving for $\omega$ to lowest order in $\mu - \mu_0$ gives the following result for the imaginary part,
\begin{equation}
\text{Im} \, \omega = \frac{1}{2} c ( \mu - \mu_0 )^2.
\end{equation}
where
\begin{equation}
c = \frac{2 \overline{N} D^3 M^2 \mu_0 |C|}{[ ( \overline{N} D )^2 + C^2 ]^2}.
\end{equation}
Since, $\text{Im} \, \omega > 0$ \emph{both} to the right \emph{and} to the left of $\mu = \mu_0$, there is potential for superradiant instability even in the region $\mu < \mu_0$, but the lack of quasibound states in that region prevents the instability from developing. In the region $\mu > \mu_0$, however, there are quasibound states and so superradiant instability does occur, up to some cut-off $\mu$, which we denote by $\mu_1$, where $\text{Im} \, \omega$ once again goes to zero and switches sign from positive to negative, as shown in fig. \ref{fig:imaginary_omega}.  

We can determine the point $\mu_1$ where $\text{Im} \, \omega$ becomes negative perturbatively in the smallness of $\mu - \omega$. To first order (i.e., keeping only up to linear order in $\mu - \omega$), this point is
\begin{equation} \label{eq:mu11}
\mu_{1}^{(1)} = \frac{A}{B} = \frac{am + qQ}{2M^2 - Q^2}.
\end{equation}
To second order (i.e., keeping up to quadratic order in $\mu - \omega$), this point is found to be
\begin{equation} \label{eq:mu12}
\mu_{1}^{(2)} = \mu_{1}^{(1)} \biggl( 1 + \frac{M^2 C^2}{2 \Nbar^2 B^2 + M^2 C \widetilde{C}} \biggr),
\end{equation}
where
\begin{equation}
\widetilde{C} = 3 am - \mu_0 ( Q^2 - 5M^2 ).
\end{equation}
Surprisingly, it is possible to solve for $\mu_1$, the boundary of the superradiant region, exactly. The result is
\begin{equation} \label{eq:mu1}
\begin{aligned}
\mu_{1} &= \biggl[ \mu_{1}^{(1)} \bigl( 2 \mu_{1}^{(1)} - \mu_0 \bigr) + \frac{1}{2} \biggl( \frac{\Nbar}{M} \biggr)^2 \\
&\quad - \frac{\Nbar}{M} \biggl[ \mu_{1}^{(1)} \bigl( \mu_{1}^{(1)} - \mu_0 \bigr) + \frac{1}{4} \biggl( \frac{\Nbar}{M} \biggr)^2 \biggr]^{\frac{1}{2}} \biggr]^{\frac{1}{2}}.
\end{aligned}
\end{equation}
We compare these results with the numerical work of Furuhashi and Nambu \cite{furuhashi2004instability} by producing a plot of $q Q$ versus $\mu M$ showing the shaded region where there are both quasibound states and superradiance. This is done numerically in fig. 3 of \cite{furuhashi2004instability}, but we can plot the boundary by eye since it is approximately linear. We can also plot our expressions, $\mu_{1}^{(1)}$ to first order, $\mu_{1}^{(2)}$ to second order, and the exact result, $\mu_1$ to compare. We include several data points along the boundary, to verify that our exact analytic expression \eqref{eq:mu1} is correct. It also proves to be extremely well approximated by our second-order $\mu_{1}^{(2)}$ analytic result. These results are shown in fig. \ref{fig:FNcompare}. We have set $N=1$ since the mapping to the principal and angular quantum numbers is $N = n + \ell$ and $n = 0$ and $\ell = 1$ in \cite{furuhashi2004instability}. Similar plots in the literature, using slightly different parameters, could also be compared to our results (e.g., the continued fractions results of \cite{Huang:2018qdl}).

%


\subsection{The Analytics vs. Numerics Discrepancy}

\begin{figure*}[t]
\includegraphics[width = 0.95 \textwidth]{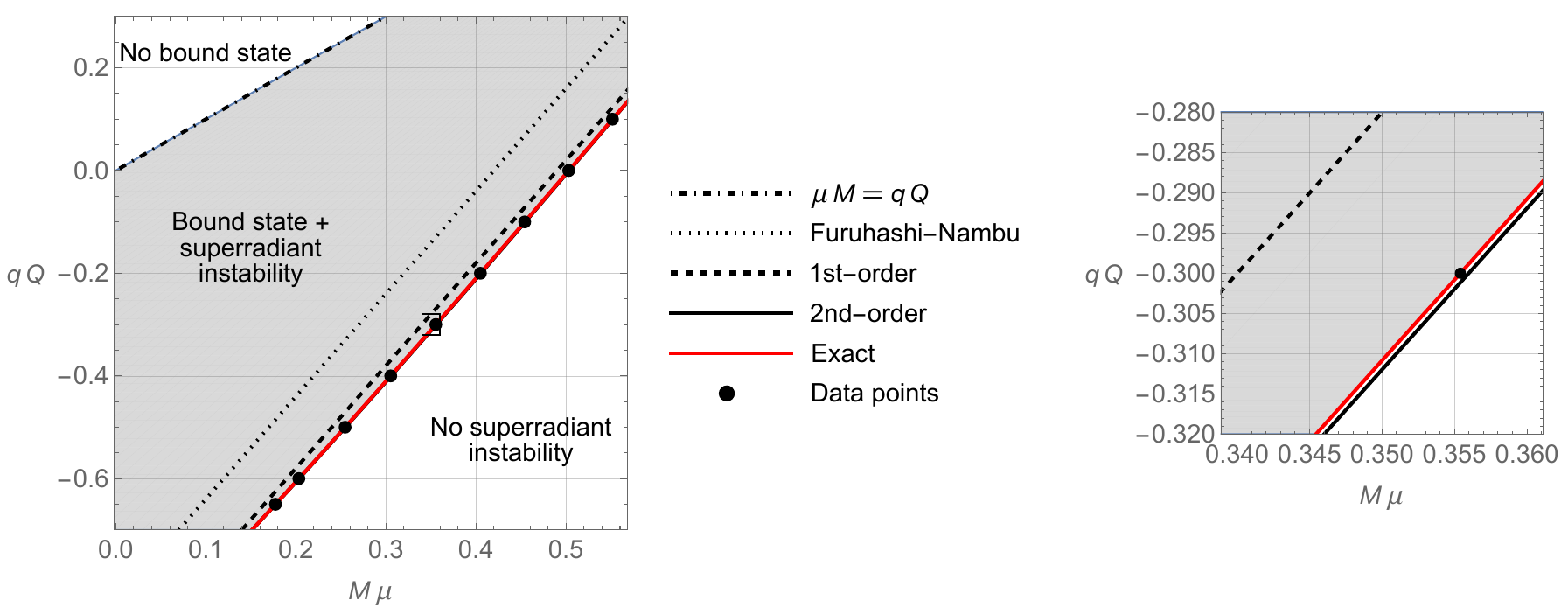}
\caption{Plot showing the regions in the $\mu M$ and $qQ$ plane (parameter values $a = 0.98 M$, $Q = 0.01 M$, $m=1$, $N=1$) where there are quasibound states and superradiance (shaded gray). The dotdashed line corresponds to $\mu M = q Q$ and is the boundary of the region with no quasibound states. The boundary of the region with no superradiance is the dotted line according to the numerical analysis of Furuhashi and Nambu \cite{furuhashi2004instability}, the dashed line according to our first-order analytic result \eqref{eq:mu11}, the solid black line according to our second-order analytic result \eqref{eq:mu12}, and the solid red line according to our exact analytic result \eqref{eq:mu1}. The circle points are exact results from solving the resonance equation of Vieira, Bezerra, and Muniz \cite{vieira2022instability}. The plot on the right zooms in on the square region outlined in the plot on the left in order to show the difference between the 2nd-order result and the exact result.}
\label{fig:FNcompare}
\end{figure*}

In \cite{vieira2022instability}, the authors claim that their plot of $\text{Im} \, \omega$ in their fig. 3 shows that the resonant frequencies ``have a behavior which is similar to the results presented'' in fig. 6 of \cite{furuhashi2004instability}. We have three qualifications to this statement:
\begin{enumerate}[1.]
\item The statement compares apples to oranges: the parameters in fig. 3 of \cite{vieira2022instability} are $m = -1$, $q=1$, $a = 0.99 M$, and $Q=0.01 M$, whereas those in fig. 6 of \cite{furuhashi2004instability} are $m=1$, $a = 0.98 M$, and $Q = 0.01M$ and the parameter $q$ is the one being tuned.

\item Fig. 3 in \cite{vieira2022instability} does not depict the behavior of $\text{Im} \, \omega$ where it vanishes near the point $\mu = \mu_0 = qQ/ M$. This is because $\mu_0 = 0.01$ in this case and one would need to zoom in significantly towards the origin of these plots to show this. The corresponding (dashed) curve in fig. 6 of \cite{furuhashi2004instability} uses $q=0$ instead and so $\mu_0 = 0$ in this case.

\item Most importantly, there is a \emph{quantative} difference between where the numerics of Furuhashi and Nambu place the boundary line beyond which superradiance stops and where our analytic analysis of the VBK resonance equation to first- and second-order places this boundary line, as can be seen in fig. \ref{fig:FNcompare}.
\end{enumerate}

This quantitative tension between the exact analytic results of \cite{vieira2022instability} and the numerical results of \cite{furuhashi2004instability} arises for the following reason. The radial equation that Furuhashi and Nambu analyze numerically contains an effective potential that takes the form of the one that comes from the hydrogenic approximation. By this, we mean that they take the separation constant that comes from the method of separation of variables to have the form $\lambda = \ell ( \ell + 1 )$, where $\ell$ is an angular momentum quantum number whose relation to the ``magnetic'' quantum number $m$ is the usual one that is appropriate for the hydrogen atom: $\ell = 0, 1, 2, \ldots$ and $- \ell \leq m \leq \ell$. In making their numerical plots, Furuhashi and Nambu take $\ell = 1$ (and $m=1$), for example. These, however, are not the actual solutions of the angular equation of interest \eqref{eq:KGangle}. The degree to which they are good approximate solutions is the degree to which $\omega \approx \mu$ and/or $\mu M , \omega M \ll 1$. From fig. \ref{fig:FNcompare}, it is clear that the boundary line where superradiance stops occurs at values of $\mu M$ that are not $\ll 1$. Furthermore, our first-order \eqref{eq:mu11} and second-order \eqref{eq:mu12} analytic boundary lines plotted in fig. \ref{fig:FNcompare} are first- and second-order precisely in the difference $\mu - \omega$. In this respect, the result of Furuhashi and Nambu may be thought of as the ``zeroth-order'' counterpart to our first- and second-order results. 

It is a remarkable fact that the characteristic resonance equation for $\omega$ is independent of the separation constant $\lambda$. The authors of \cite{vieira2022instability} point this out explicitly. It is what allows them to write down the resonance equation without having to determine the allowed values of the separation constant by solving the angular equation first. Therefore, the numerics may get different boundary lines for different values of $\ell$ and they almost certainly will. Nevertheless, the true boundary line, as well as our first- and second-order approximations of it, do not depend on $\lambda$ at all and, therefore, do not depend on how well $\lambda$ is approximated by the form $\lambda = \ell ( \ell + 1)$. The dependence on $\ell$ is instead \emph{implicit} through the so-called ``overtone number'' $N$ (or $\overline{N} = N+1$) in that, at least in the hydrogenic limit, $N$ would be related to the principal quantum number $n$ and the angular momentum quantum number $\ell$ via $N = n + \ell$.

%


\section{Conclusion and Outlook}
\label{sec:6}

In this paper we used the Vieira-Bezerra-Kokkotas (VBK) method to specify the regions of black hole superradiant instability as a function of the black hole parameters (spin $a$ and charge $Q$) and scalar field mass $\mu$. This exact method allows us to obtain analytic control over the parameter space of quasibound and superradiant modes without resorting to the hydrogenic approximation or numerics. 

We solved for the quasibound states and the superradiant states and plotted their respective regions in the parameter space. The overlap regions determine the parameters in which superradiant instability will occur. We found that for very small values of $\mu$, we have superradiant instability for black holes that have a very small charge $Q < \mu$ (normalized by $M$ and for the particular value $qM = 1$). Likewise, we also determined that for black holes that have very small spin values $a$, the overlap region vanishes. This is especially apparent if we take larger values of $\mu$. To compensate, the black hole must then be highly charged and closer to the extremal case in order to potentially be unstable to superradiance. Essentially,
\begin{enumerate}[$\bullet$]
    \item For small values of the scalar field mass $\mu M \ll 1$, black holes that are closer to being ``Kerr-like" than ``RN-like" are more susceptible to superradiant instability.

    \item For large values of the scalar field mass $\mu M \lesssim 1$, Kerr-Newman black holes that are highly charged, have non-zero spin $a$, and are close to extremal are the ones which are susceptible to superradiant instability.

    \item For in-between values of $\mu$, refer to fig. \ref{subfig:instability}.
\end{enumerate}
Thus, we have provided a unified and exact characterization of the region of superradiant instability of Kerr-Newman black holes in the presence of a massive real scalar field. These results extend the standard results in the literature, which were mostly confined to small scalar field mass $\mu M \ll 1$ or Kerr black holes.

We also performed an expansion of the characteristic resonance equation \eqref{eq:resoeq1} near the point $\mu = \mu_0$, which is the point where $\omega = \mu$. The parameter $\text{Re} \bigl[ \sqrt{\mu^2 - \omega^2} \bigr]$ is positive for $\mu > \mu_0$, meaning that the scalar field decays towards infinite radial distance from the black hole and is thus quasibound, and is negative for $\mu < \mu_0$. This tells us that quasibound states will only form for scalar fields with $\mu > \mu_0$. We also checked the region around $\mu_0$ and determined if these parameters satisfy $\text{Im}\,\omega > 0$, thus allowing the scalar field to grow exponentially over time. We discovered that in the regions $\mu < \mu_0$ and $\mu > \mu_0$, $\text{Im} \, \omega > 0$, hence, in both regions the scalar field mode can grow. However, since the region $\mu < \mu_0$ does not satisfy the quasibound state condition, black hole instability does not develop there. Thus, only in the region $\mu > \mu_0$ do we expect to have instabilities due to superradiance.    

We derived analytic expressions for the boundary line beyond which $\text{Im}\, \omega$ becomes negative. Beyond this line, the scalar field no longer grows over time preventing the development of a superradiant instability. These expressions are $\mu_{1}^{(1)}$ \eqref{eq:mu11} and $\mu_{1}^{(2)}$ \eqref{eq:mu12}, to linear and quadratic order in $\mu - \omega$, respectively. We can also give the exact analytic expression \eqref{eq:mu1} for this boundary line. This is compared with several data points of the \emph{exact} result derived from the plot of the exact solution for $\text{Im}\, \omega$ as a function of $\mu$ and fig. \ref{fig:FNcompare} shows good agreement between the exact result and the second-order result $\mu_{1}^{(2)}$. There is a clear difference between the exact result \eqref{eq:mu1} and the numerical results of Furuhashi and Nambu \cite{furuhashi2004instability}. We attribute the discrepancy to the approximation $\omega \approx \mu$ used in the numerical analysis and therefore our first- and second-order results can be thought of as corrections to the numerics. Many works in the literature use the numerical results of Furuhashi and Nambu to benchmark various analytic methods. Therefore, these should be re-evaluated. For example, sizable discrepancies between analytic results for the maximum growth rate and numerical results of Furuhashi and Nambu were reported in \cite{Bao:2023xna}. It would be useful to compare with our exact results instead to check if these discrepancies persist. 

As possible extensions of this work, it would be interesting to study other black hole spacetimes outside the Kerr-Newman family, such as Kerr-Newman-Kasuya (dyonic) black holes. This would build on existing work on the superradiant instability of magnetically charged black holes in the presence of an electrically charged massive scalar field, where new phenomenology and structural differences in the instability apparently arise \cite{Pereniguez:2024fkn}. It would also be interesting to study spacetimes with different boundary or asymptotic conditions (e.g., anti-de Sitter spacetime). We can also study fields with higher spin (e.g., massive vector fields). There are also opportunities in further numerical analyses of the aforementioned systems and potentially including backreaction. Finally, it would be interesting to study systems of astrophysical interest, including binary systems, and making contact with gravitational waves.


\acknowledgments

K.T.G. would like to acknowledge financial support from the National Institute of Physics, University of the Philippines, Diliman Research Grant Project ``Superradiance and Quasi-Bound States in the Parameter Space of the Kerr-Newman Black Hole.'' The authors also acknowledge the Office of the Chancellor of the University of the Philippines Diliman, through the Office of the Vice Chancellor for Research and Development, for funding support through the Outright Research Grant (262608 ORG).


\bibliography{KNSI}

@article{cole2023distinguishing,
    author = "Cole, Philippa S. and Bertone, Gianfranco and Coogan, Adam and Gaggero, Daniele and Karydas, Theophanes and Kavanagh, Bradley J. and Spieksma, Thomas F. M. and Tomaselli, Giovanni Maria",
    title = "{Distinguishing environmental effects on binary black hole gravitational waveforms}",
    eprint = "2211.01362",
    archivePrefix = "arXiv",
    primaryClass = "gr-qc",
    doi = "10.1038/s41550-023-01990-2",
    journal = "Nature Astron.",
    volume = "7",
    number = "8",
    pages = "943--950",
    year = "2023"
}

@article{cardoso2022black,
    author = "Cardoso, Vitor and Destounis, Kyriakos and Duque, Francisco and Macedo, Rodrigo Panosso and Maselli, Andrea",
    title = "{Black holes in galaxies: Environmental impact on gravitational-wave generation and propagation}",
    eprint = "2109.00005",
    archivePrefix = "arXiv",
    primaryClass = "gr-qc",
    doi = "10.1103/PhysRevD.105.L061501",
    journal = "Phys. Rev. D",
    volume = "105",
    number = "6",
    pages = "L061501",
    year = "2022"
}

@article{sarmah2025effects,
    author = "Sarmah, Priyanka and Verma, Himanshu and Cheung, Kingman and Silk, Joseph",
    title = "{Effects of superradiance in active galactic nuclei}",
    eprint = "2404.09955",
    archivePrefix = "arXiv",
    primaryClass = "astro-ph.HE",
    doi = "10.1093/mnras/staf326",
    journal = "Mon. Not. Roy. Astron. Soc.",
    volume = "538",
    number = "2",
    pages = "943--962",
    year = "2025"
}

@incollection{lupsasca2024beginner,
    author = "Lupsasca, Alexandru and Mayerson, Daniel R. and Ripperda, Bart and Staelens, Seppe",
    editor = "Bambi, Cosimo and Cardenas-Avendano, Alejandro",
    title = "{A Beginner{\textquoteright}s Guide to~Black Hole Imaging and~Associated Tests of~General Relativity}",
    booktitle = "{Recent Progress on Gravity Tests. Challenges and Future Perspectives}",
    eprint = "2402.01290",
    archivePrefix = "arXiv",
    primaryClass = "gr-qc",
    pages = "183--237",
    year = "2024",
    publisher={\href{https://link.springer.com/chapter/10.1007/978-981-97-2871-8_6}{Springer}}
}

@article{oei2024black,
    author = "Oei, Martijn S.S.L. and Hardcastle, Martin J and Timmerman, Roland and others",
    title = "{Black hole jets on the scale of the cosmic web}",
    eprint = "2411.08630",
    archivePrefix = "arXiv",
    primaryClass = "astro-ph.HE",
    doi = "10.1038/s41586-024-07879-y",
    journal = "Nature",
    volume = "633",
    number = "8030",
    pages = "537--541",
    year = "2024"
}

@article{brito2025black,
    author = "Brito, Richard",
    title = "{Black holes as laboratories: searching for ultralight fields}",
    doi = "10.1007/s10714-025-03376-3",
    journal = "Gen. Rel. Grav.",
    volume = "57",
    number = "2",
    pages = "42",
    year = "2025"
}

@inproceedings{blas2022black,
    author = "Blas, D.",
    title = "{Black hole superradiance to search for new particles}",
    booktitle = "{33rd Rencontres de Blois}: {Exploring the Dark Universe}",
    eprint = "2211.02067",
    archivePrefix = "arXiv",
    primaryClass = "hep-ph",
    month = "11",
    year = "2022"
}

@article{hui2023black,
    author = "Hui, Lam and Law, Y. T. Albert and Santoni, Luca and Sun, Guanhao and Tomaselli, Giovanni Maria and Trincherini, Enrico",
    title = "{Black hole superradiance with dark matter accretion}",
    eprint = "2208.06408",
    archivePrefix = "arXiv",
    primaryClass = "gr-qc",
    doi = "10.1103/PhysRevD.107.104018",
    journal = "Phys. Rev. D",
    volume = "107",
    number = "10",
    pages = "104018",
    year = "2023"
}

@article{brito2020superradiance,
    author = "Brito, Richard and Cardoso, Vitor and Pani, Paolo",
    title = "{Superradiance}: {New Frontiers in Black Hole
Physics}",
    eprint = "1501.06570",
    archivePrefix = "arXiv",
    primaryClass = "gr-qc",
    doi = "10.1007/978-3-319-19000-6",
    journal = "Lect. Notes Phys.",
    volume = "906",
    pages = "pp.1--237",
    year = "2015"
}

@article{bekenstein1998many,
    author = "Bekenstein, Jacob D. and Schiffer, Marcelo",
    title = "{The Many faces of superradiance}",
    eprint = "gr-qc/9803033",
    archivePrefix = "arXiv",
    doi = "10.1103/PhysRevD.58.064014",
    journal = "Phys. Rev. D",
    volume = "58",
    pages = "064014",
    year = "1998"
}

@article{lukin2022optical,
    author = "Lukin, Daniil M. and Guidry, Melissa A. and Yang, Joshua and Ghezellou, Misagh and Mishra, Sattwik Deb and Abe, Hiroshi and Ohshima, Takeshi and Ul-Hassan, Jawad and Vu{\v{c}}kovi{\'c}, Jelena",
    title = "{Optical superradiance of a pair of color centers in an integrated silicon-carbide-on-insulator microresonator}",
    eprint = "2202.04845",
    archivePrefix = "arXiv",
    primaryClass = "quant-ph",
    month = "2",
    year = "2022"
}

@article{sierra2022dicke,
    author = "Sierra, Eric and Masson, Stuart J. and Asenjo-Garcia, Ana",
    title = "{Dicke Superradiance in Ordered Lattices: Dimensionality Matters}",
    eprint = "2110.08380",
    archivePrefix = "arXiv",
    primaryClass = "quant-ph",
    doi = "10.1103/PhysRevResearch.4.023207",
    journal = "Phys. Rev. Res.",
    volume = "4",
    number = "2",
    pages = "023207",
    year = "2022"
}

@article{stransky2019superradiance,
    author = "Str{\'a}nsk{\'y}, Pavel and Cejnar, Pavel",
    title = "{Superradiance in finite quantum systems randomly coupled to continuum}",
    eprint = "1907.11576",
    archivePrefix = "arXiv",
    primaryClass = "quant-ph",
    doi = "10.1103/PhysRevE.100.042119",
    journal = "Phys. Rev. E",
    volume = "100",
    pages = "042119",
    year = "2019"
}

@article{eleuch2014open,
    author = "Eleuch, Hichem and Rotter, Ingrid",
    title = "{Open quantum systems and Dicke superradiance}",
    journal={Eur. Phys. J. D},
    volume={68},
    number={3},
    pages={74},
    year={2014},
    publisher={Springer},
    doi = "10.1140/epjd/e2014-40780-8",
    eprint = "1305.2762",
    archivePrefix = "arXiv",
    primaryClass = "quant-ph"
}

@article{blackholebomb,
    author = "Press, William H. and Teukolsky, Saul A.",
    title = "{Floating Orbits, Superradiant Scattering and the Black-hole Bomb}",
    doi = "10.1038/238211a0",
    journal = "Nature",
    volume = "238",
    pages = "211-212",
    year = "1972"
}

@article{cardoso2004black,
    author = "Cardoso, Vitor and Dias, Oscar J. C. and Lemos, Jose P. S. and Yoshida, Shijun",
    title = "{The Black hole bomb and superradiant instabilities}",
    eprint = "hep-th/0404096",
    archivePrefix = "arXiv",
    doi = "10.1103/PhysRevD.70.049903",
    journal = "Phys. Rev. D",
    volume = "70",
    pages = "044039",
    year = "2004",
    note = "[Erratum: Phys.Rev.D 70, 049903 (2004)]"
}

@article{Damour:1976kh,
    author = "Damour, T. and Deruelle, N. and Ruffini, R.",
    title = "{On Quantum Resonances in Stationary Geometries}",
    doi = "10.1007/BF02725534",
    journal = "Lett. Nuovo Cim.",
    volume = "15",
    pages = "257--262",
    year = "1976"
}

@article{ZOUROS1979139,
title = {Instabilities of massive scalar perturbations of a rotating black hole},
author = {Theodoros J.M Zouros and Douglas M Eardley},
journal = {Annals of Physics},
volume = {118},
number = {1},
pages = {139-155},
year = {1979},
issn = {0003-4916},
doi = {https://doi.org/10.1016/0003-4916(79)90237-9},
url = {https://www.sciencedirect.com/science/article/pii/0003491679902379}
}

@article{press1972floating,
	title = {Floating Orbits, Superradiant Scattering and the Black-hole Bomb},
	author={Press, William H and Teukolsky, Saul A},
	date = {1972/07/01},
	doi = {10.1038/238211a0},
	id = {PRESS1972},
	isbn = {1476-4687},
	journal = {Nature},
	number = {5361},
	pages = {211--212},
	url = {https://doi.org/10.1038/238211a0},
	volume = {238},
	year = {1972}
	}

@article{baumann2019spectra,
    author = "Baumann, Daniel and Chia, Horng Sheng and Stout, John and ter Haar, Lotte",
    title = "{The Spectra of Gravitational Atoms}",
    eprint = "1908.10370",
    archivePrefix = "arXiv",
    primaryClass = "gr-qc",
    doi = "10.1088/1475-7516/2019/12/006",
    journal = "JCAP",
    volume = "12",
    pages = "006",
    year = "2019"
}

@article{dias2023black,
    author = "Dias, Oscar J. C. and Lingetti, Giuseppe and Pani, Paolo and Santos, Jorge E.",
    title = "{Black hole superradiant instability for massive spin-2 fields}",
    eprint = "2304.01265",
    archivePrefix = "arXiv",
    primaryClass = "gr-qc",
    doi = "10.1103/PhysRevD.108.L041502",
    journal = "Phys. Rev. D",
    volume = "108",
    number = "4",
    pages = "L041502",
    year = "2023"
}

@article{tomaselli2024legacy,
    author = "Tomaselli, Giovanni Maria and Spieksma, Thomas F. M. and Bertone, Gianfranco",
    title = "{Legacy of Boson Clouds on Black Hole Binaries}",
    eprint = "2407.12908",
    archivePrefix = "arXiv",
    primaryClass = "gr-qc",
    doi = "10.1103/PhysRevLett.133.121402",
    journal = "Phys. Rev. Lett.",
    volume = "133",
    number = "12",
    pages = "121402",
    year = "2024"
}

@article{tomaselli2024resonant,
    author = "Tomaselli, Giovanni Maria and Spieksma, Thomas F. M. and Bertone, Gianfranco",
    title = "{Resonant history of gravitational atoms in black hole binaries}",
    eprint = "2403.03147",
    archivePrefix = "arXiv",
    primaryClass = "gr-qc",
    doi = "10.1103/PhysRevD.110.064048",
    journal = "Phys. Rev. D",
    volume = "110",
    number = "6",
    pages = "064048",
    year = "2024"
}

@article{Baryakhtar:2020gao,
    author = "Baryakhtar, Masha and Galanis, Marios and Lasenby, Robert and Simon, Olivier",
    title = "{Black hole superradiance of self-interacting scalar fields}",
    eprint = "2011.11646",
    archivePrefix = "arXiv",
    primaryClass = "hep-ph",
    doi = "10.1103/PhysRevD.103.095019",
    journal = "Phys. Rev. D",
    volume = "103",
    number = "9",
    pages = "095019",
    year = "2021"
}

@article{baryakhtar2017black,
    author = "Baryakhtar, Masha and Lasenby, Robert and Teo, Mae",
    title = "{Black Hole Superradiance Signatures of Ultralight Vectors}",
    eprint = "1704.05081",
    archivePrefix = "arXiv",
    primaryClass = "hep-ph",
    doi = "10.1103/PhysRevD.96.035019",
    journal = "Phys. Rev. D",
    volume = "96",
    number = "3",
    pages = "035019",
    year = "2017"
}

@article{vicente2018penrose,
    author = "Vicente, Rodrigo and Cardoso, Vitor and Lopes, Jorge C.",
    title = "{Penrose process, superradiance, and ergoregion instabilities}",
    eprint = "1803.08060",
    archivePrefix = "arXiv",
    primaryClass = "gr-qc",
    doi = "10.1103/PhysRevD.97.084032",
    journal = "Phys. Rev. D",
    volume = "97",
    number = "8",
    pages = "084032",
    year = "2018"
}

@article{starobinskii1973amplification,
    author = "Starobinskii, A. A.",
    title = "{Amplification of waves during reflection from a rotating ``black hole''}",
    journal = "Sov. Phys. JETP",
    volume = "37",
    number = "1",
    pages = "28--32",
    year = "1973"
}

@article{detweiler1980klein,
  title = {Klein-Gordon equation and rotating black holes},
  author = {Detweiler, Steven},
  journal = {Phys. Rev. D},
  volume = {22},
  issue = {10},
  pages = {2323--2326},
  numpages = {0},
  year = {1980},
  month = {Nov},
  publisher = {American Physical Society},
  doi = {10.1103/PhysRevD.22.2323},
  url = {https://link.aps.org/doi/10.1103/PhysRevD.22.2323}
}

@article{Konoplya:2019hlu,
    author = "Konoplya, R. A. and Zhidenko, A. and Zinhailo, A. F.",
    title = "{Higher order WKB formula for quasinormal modes and grey-body factors: recipes for quick and accurate calculations}",
    eprint = "1904.10333",
    archivePrefix = "arXiv",
    primaryClass = "gr-qc",
    doi = "10.1088/1361-6382/ab2e25",
    journal = "Class. Quant. Grav.",
    volume = "36",
    pages = "155002",
    year = "2019"
}

@article{Konoplya:2023moy,
    author = "Konoplya, R. A. and Zhidenko, A.",
    title = "{Analytic expressions for quasinormal modes and grey-body factors in the eikonal limit and beyond}",
    eprint = "2309.02560",
    archivePrefix = "arXiv",
    primaryClass = "gr-qc",
    doi = "10.1088/1361-6382/ad0a52",
    journal = "Class. Quant. Grav.",
    volume = "40",
    number = "24",
    pages = "245005",
    year = "2023"
}

@article{Suzuki:1998vy,
    author = "Suzuki, Hisao and Takasugi, Eiichi and Umetsu, Hiroshi",
    title = "{Perturbations of Kerr-de Sitter black hole and Heun's equations}",
    eprint = "gr-qc/9805064",
    archivePrefix = "arXiv",
    reportNumber = "EPHOU-98-005, OU-HET-296",
    doi = "10.1143/PTP.100.491",
    journal = "Prog. Theor. Phys.",
    volume = "100",
    pages = "491--505",
    year = "1998"
}

@article{Dolan:2007mj,
    author = "Dolan, Sam R.",
    title = "{Instability of the massive Klein-Gordon field on the Kerr spacetime}",
    eprint = "0705.2880",
    archivePrefix = "arXiv",
    primaryClass = "gr-qc",
    doi = "10.1103/PhysRevD.76.084001",
    journal = "Phys. Rev. D",
    volume = "76",
    pages = "084001",
    year = "2007"
}

@article{Konoplya:2024vuj,
    author = "Konoplya, R. A. and Zhidenko, A.",
    title = "{Correspondence between grey-body factors and quasinormal frequencies for rotating black holes}",
    eprint = "2408.11162",
    archivePrefix = "arXiv",
    primaryClass = "gr-qc",
    doi = "10.1016/j.physletb.2025.139288",
    journal = "Phys. Lett. B",
    volume = "861",
    pages = "139288",
    year = "2025"
}

@article{arvanitaki2015discovering,
    author = "Arvanitaki, Asimina and Baryakhtar, Masha and Huang, Xinlu",
    title = "{Discovering the QCD Axion with Black Holes and Gravitational Waves}",
    eprint = "1411.2263",
    archivePrefix = "arXiv",
    primaryClass = "hep-ph",
    doi = "10.1103/PhysRevD.91.084011",
    journal = "Phys. Rev. D",
    volume = "91",
    number = "8",
    pages = "084011",
    year = "2015"
}

@article{baumann2022ionization,
    author = "Baumann, Daniel and Bertone, Gianfranco and Stout, John and Tomaselli, Giovanni Maria",
    title = "{Ionization of gravitational atoms}",
    eprint = "2112.14777",
    archivePrefix = "arXiv",
    primaryClass = "gr-qc",
    doi = "10.1103/PhysRevD.105.115036",
    journal = "Phys. Rev. D",
    volume = "105",
    number = "11",
    pages = "115036",
    year = "2022"
}

@article{Vieira:2021xqw,
    author = "Vieira, H. S. and Kokkotas, Kostas D.",
    title = "{Quasibound states of Schwarzschild acoustic black holes}",
    eprint = "2104.03938",
    archivePrefix = "arXiv",
    primaryClass = "gr-qc",
    doi = "10.1103/PhysRevD.104.024035",
    journal = "Phys. Rev. D",
    volume = "104",
    number = "2",
    pages = "024035",
    year = "2021"
}

@article{vieira2022instability,
    author = "Vieira, H. S. and Bezerra, V. B. and Muniz, C. R.",
    title = "{Instability of the charged massive scalar field on the Kerr{\textendash}Newman black hole spacetime}",
    eprint = "2107.02562",
    archivePrefix = "arXiv",
    primaryClass = "gr-qc",
    doi = "10.1140/epjc/s10052-022-10908-7",
    journal = "Eur. Phys. J. C",
    volume = "82",
    number = "10",
    pages = "932",
    year = "2022"
}

@article{dolan2007instability,
    author = "Dolan, Sam R.",
    title = "{Instability of the massive Klein-Gordon field on the Kerr spacetime}",
    eprint = "0705.2880",
    archivePrefix = "arXiv",
    primaryClass = "gr-qc",
    doi = "10.1103/PhysRevD.76.084001",
    journal = "Phys. Rev. D",
    volume = "76",
    pages = "084001",
    year = "2007"
}

@article{east2017superradiant,
    author = "East, William E. and Pretorius, Frans",
    title = "{Superradiant Instability and Backreaction of Massive Vector Fields around Kerr Black Holes}",
    eprint = "1704.04791",
    archivePrefix = "arXiv",
    primaryClass = "gr-qc",
    doi = "10.1103/PhysRevLett.119.041101",
    journal = "Phys. Rev. Lett.",
    volume = "119",
    number = "4",
    pages = "041101",
    year = "2017"
}

@article{arvanitaki2011exploring,
    author = "Arvanitaki, Asimina and Dubovsky, Sergei",
    title = "{Exploring the String Axiverse with Precision Black Hole Physics}",
    eprint = "1004.3558",
    archivePrefix = "arXiv",
    primaryClass = "hep-th",
    doi = "10.1103/PhysRevD.83.044026",
    journal = "Phys. Rev. D",
    volume = "83",
    pages = "044026",
    year = "2011"
}

@article{yoshino2012bosenova,
    author = "Yoshino, Hirotaka and Kodama, Hideo",
    title = "{Bosenova collapse of axion cloud around a rotating black hole}",
    eprint = "1203.5070",
    archivePrefix = "arXiv",
    primaryClass = "gr-qc",
    reportNumber = "KEK-TH-1530",
    doi = "10.1143/PTP.128.153",
    journal = "Prog. Theor. Phys.",
    volume = "128",
    pages = "153--190",
    year = "2012"
}

@article{baryakhtar2021black,
    author = "Baryakhtar, Masha and Galanis, Marios and Lasenby, Robert and Simon, Olivier",
    title = "{Black hole superradiance of self-interacting scalar fields}",
    eprint = "2011.11646",
    archivePrefix = "arXiv",
    primaryClass = "hep-ph",
    doi = "10.1103/PhysRevD.103.095019",
    journal = "Phys. Rev. D",
    volume = "103",
    number = "9",
    pages = "095019",
    year = "2021"
}

@article{xie2025self,
    author = "Xie, Ning and Huang, Fa Peng",
    title = "{Self-interaction effects on the Kerr black hole superradiance and their observational implications}",
    eprint = "2503.10347",
    archivePrefix = "arXiv",
    primaryClass = "hep-ph",
    doi = "10.1103/xmhn-cpv4",
    journal = "Phys. Rev. D",
    volume = "112",
    number = "5",
    pages = "055028",
    year = "2025"
}

@article{Myung:2022biw,
    author = "Myung, Yun Soo",
    title = "{Conditions for superradiant instability of the Kerr-Newman black holes}",
    eprint = "2204.06750",
    archivePrefix = "arXiv",
    primaryClass = "gr-qc",
    doi = "10.1103/PhysRevD.105.124015",
    journal = "Phys. Rev. D",
    volume = "105",
    number = "12",
    pages = "124015",
    year = "2022"
}

@article{furuhashi2004instability,
    author = "Furuhashi, Hironobu and Nambu, Yasusada",
    title = "{Instability of massive scalar fields in Kerr-Newman space-time}",
    eprint = "gr-qc/0402037",
    archivePrefix = "arXiv",
    reportNumber = "DPNU-03-28",
    doi = "10.1143/PTP.112.983",
    journal = "Prog. Theor. Phys.",
    volume = "112",
    pages = "983--995",
    year = "2004"
}

@article{adamo2014kerr,
    author = "Adamo, Tim and Newman, E. T.",
    title = "{The Kerr-Newman metric: A Review}",
    eprint = "1410.6626",
    archivePrefix = "arXiv",
    primaryClass = "gr-qc",
    doi = "10.4249/scholarpedia.31791",
    journal = "Scholarpedia",
    volume = "9",
    pages = "31791",
    year = "2014"
}

@article{Berti:2005gp,
    author = "Berti, Emanuele and Cardoso, Vitor and Casals, Marc",
    title = "{Eigenvalues and eigenfunctions of spin-weighted spheroidal harmonics in four and higher dimensions}",
    eprint = "gr-qc/0511111",
    archivePrefix = "arXiv",
    doi = "10.1103/PhysRevD.73.109902",
    journal = "Phys. Rev. D",
    volume = "73",
    pages = "024013",
    year = "2006",
    note = "[Erratum: Phys.Rev.D 73, 109902 (2006)]"
}

@article{moncrief1974stability,
    author = "Moncrief, Vincent",
    title = "{Stability of Reissner-Nordstr\"om black holes}",
    doi = "10.1103/PhysRevD.10.1057",
    journal = "Phys. Rev. D",
    volume = "10",
    pages = "1057--1059",
    year = "1974"
}

@article{hod2013no,
  title={No-bomb theorem for charged {{R}eissner}-{{N}ordstr{\"o}m} black holes},
  author={Hod, Shahar},
  journal={Phys Lett B},
  volume={718},
  number={4-5},
  pages={1489--1492},
  year={2013},
  doi = {https://doi.org/10.1016/j.physletb.2012.12.013},
   eprint = "1304.6474",
    archivePrefix = "arXiv",
    primaryClass = "gr-qc"
}

@article{hod2015stability,
    author = "Hod, Shahar",
    title = {{Stability of highly-charged Reissner-Nordstr{\"o}m black holes to charged scalar perturbations}},
    eprint = "1504.00009",
    archivePrefix = "arXiv",
    primaryClass = "gr-qc",
    doi = "10.1103/PhysRevD.91.044047",
    journal = "Phys. Rev. D",
    volume = "91",
    number = "4",
    pages = "044047",
    year = "2015"
}

@article{Huang:2018qdl,
    author = "Huang, Yang and Liu, Dao-Jun and Zhai, Xiang-hua and Li, Xin-zhou",
    title = "{Instability for massive scalar fields in Kerr-Newman spacetime}",
    eprint = "1807.06263",
    archivePrefix = "arXiv",
    primaryClass = "gr-qc",
    doi = "10.1103/PhysRevD.98.025021",
    journal = "Phys. Rev. D",
    volume = "98",
    number = "2",
    pages = "025021",
    year = "2018"
}

@article{Bao:2023xna,
    author = "Bao, Shou-Shan and Xu, Qi-Xuan and Zhang, Hong",
    title = "{Next-to-leading-order solution to Kerr-Newman black hole superradiance}",
    eprint = "2301.05317",
    archivePrefix = "arXiv",
    primaryClass = "gr-qc",
    doi = "10.1103/PhysRevD.107.064037",
    journal = "Phys. Rev. D",
    volume = "107",
    number = "6",
    pages = "064037",
    year = "2023"
}

@article{Pereniguez:2024fkn,
    author = "Pere{\~n}iguez, David and de Amicis, Marina and Brito, Richard and Panosso Macedo, Rodrigo",
    title = "{Superradiant Instability of Magnetic Black Holes}",
    eprint = "2402.05178",
    archivePrefix = "arXiv",
    primaryClass = "gr-qc",
    doi = "10.1103/PhysRevD.110.104001",
    journal = "Phys. Rev. D",
    volume = "110",
    pages = "104001",
    year = "2024"
}

\end{document}